\documentstyle[multicol,aps,psfig]{revtex}
\renewcommand{\narrowtext}{\begin{multicols}{2}
\global\columnwidth20.5pc\noindent}
\renewcommand{\widetext}{\end{multicols}
\global\columnwidth42.5pc}
\begin{document}
\draft
\preprint{18 February 2002}
\title{Intrinsic double-peak structure of the specific heat in
       low-dimensional quantum ferrimagnets}
\author{Takashi Nakanishi and Shoji Yamamoto}
\address{Department of Physics, Okayama University,
         Tsushima, Okayama 700-8530, Japan}
%\date{Received \hspace{4cm}}
\date{18 February 2002}
\maketitle
\begin{abstract}
Motivated by recent magnetic measurements on
$A_3\mbox{Cu}_3(\mbox{PO}_4)_4$ ($A=\mbox{Ca},\mbox{Sr}$) and
$\mbox{Cu}(3\mbox{-Clpy})_2(\mbox{N}_3)_2$
($3\mbox{-Clpy}=3\mbox{-Chloropyridine}$),
both of which behave like one-dimensional ferrimagnets, we
extensively investigate the ferrimagnetic specific heat with
particular emphasis on its double-peak structure.
Developing a modified spin-wave theory, we reveal that ferromagnetic
and antiferromagnetic dual features of ferrimagnets may potentially
induce an extra low-temperature peak as well as a Schottky-type peak
at mid temperatures in the specific heat.
\end{abstract}
\narrowtext

   Recent progress on the theoretical understanding of
low-dimensional (low-D) quantum ferrimagnets deserves special mention
in their long history of research.
A minimum of the susceptibility ($\chi$)-temperature ($T$) product
has been known as typical of 1-D ferrimagnets \cite{K89}.
Although the $T^{-1}$-diverging $\chi T$ at low temperatures is
reminiscent of the ferromagnetic susceptibility, it turns into the
high-temperature paramagnetic behavior showing the antiferromagnetic
increase.
Recently an explicit sum rule \cite{Y1024} for the ferrimagnetic
susceptibility has been found:
Spin-$(S,s)$ ferrimagnetic chains behave similar to combinations of
spin-$(S-s)$ ferromagnetic and spin-$(2s)$ antiferromagnetic chains
provided $S=2s$.
An epochal argument \cite{O1984} on the ground-state magnetization
curves of low-D quantum magnets stimulated broad interest
in ferrimagnetic chains \cite{K1762,S4053} and ladders \cite{L11725}
in a field.
Spin-$(S,s)$ ferrimagnetic chains were found to exhibit $2s$
magnetization plateaus without any bond alternation \cite{Y3795}.
The discovery of a metal-oxide Haldane-gap antiferromagnet
Y$_2$BaNiO$_5$ \cite{D409,D1857} and its rare-earth derivatives
$R_2$BaNiO$_5$ \cite{Z6437,Y11516} directed our attention to
2-D mixed-spin magnets.
Their magnetic double structure, that is, the coexistence of
gapless and gapped excitations, was well interpreted from the point
of view of coupled ferrimagnetic chains \cite{T15189}.
Nuclear-magnetic-resonance measurements \cite{F433} on an ordered
bimetallic chain compound
NiCu(C$_7$H$_6$N$_2$O$_6$)(H$_2$O)$_3$$\cdot$2H$_2$O
revealed a unique field dependence of the relaxation rate,
$T_1^{-1}\propto H^{-1/2}$, which was found to be indirect
observation of ferrimagnetic dispersion relations \cite{Y842}.

\begin{figure}
\centerline
{\mbox{\psfig{figure=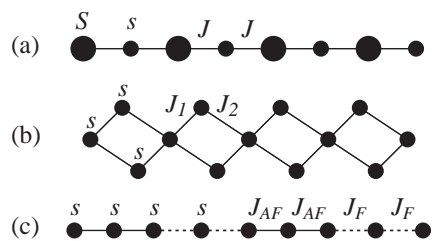,width=56mm,angle=0}}}
\vspace*{4mm}
\caption{Schematic representations of the bimetallic chain compounds
         (a), the trimeric intertwining double-chain compounds (b),
         and the tetrameric bond-alternating chain compounds (c), 
         which are described by the Hamiltonians (1a), (1b), and (1c),
         respectively, where smaller and larger bullet symbols denote
         spins $s=\frac{1}{2}$ and $S>\frac{1}{2}$, while solid and
         dashed segments mean antiferromagnetic and ferromagnetic
         exchange couplings between them, respectively.}
\label{F:illust}
\end{figure}

   In this article, we discuss another hot topic on the ferrimagnetic
specific heat ($C$).
Intertwining double-chain ferrimagnets
$A_3\mbox{Cu}_3(\mbox{PO}_4)_4$ ($A=\mbox{Ca},\mbox{Sr}$) \cite{D83}
and a ferromagnetic-ferromagnetic-antiferromagnetic-antiferromagnetic
bond-alternating chain ferrimagnet
$\mbox{Cu}(3\mbox{-Clpy})_2(\mbox{N}_3)_2$
($3\mbox{-Clpy}=3\mbox{-Chloropyridine}$) \cite{H}, which are
illustrated in Figs. \ref{F:illust}(b) and \ref{F:illust}(c),
respectively, were both reported to exhibit a unique temperature
dependence of the specific heat: A minimum at low temperatures and
then a noticeable increase toward a Schottky-type maximum.
Ferrimagnets generally possess a ground state of macroscopically
degenerate multiplet and therefore an applied field may induce a
double-peaked specific heat \cite{M5908}.
However, such an {\it extrinsic} mechanism should be distinguished
from the {\it intrinsic} thermodynamics.
Magnetic measurements \cite{D83,A186,H30} were also performed for
these materials in an attempt to evaluate the exchange couplings but
no parametrization reported so far is so consistent as to fit the
magnetization, susceptibility, and specific heat consistently.
Thus motivated, we calculate the specific heat for a wide class of
1-D ferrimagnets and reveal its intrinsic and generic features.
\vspace*{-5mm}

   The simplest quantum ferrimagnet in one dimension consists of two
kinds of spins $S$ and $s$ alternating on a ring with
antiferromagnetic exchange coupling between nearest neighbors and a
series of such family compounds were extensively synthesized by
Verdaguer, Kahn, and their coworkers \cite{K89}.
One of their works is bimetallic chains of general formula
$A$Cu(pbaOH)(H$_2$O)$_3$$\cdot$$n$H$_2$O
($A=\mbox{Ni},\mbox{Co},\mbox{Fe},\mbox{Mn}$;
 $\mbox{pbaOH}=\mbox{2-hydroxy-1,3-propylenebis(oxamato)}$)
\cite{K782}, which are illustrated in Fig. \ref{F:illust}(a).
We first consider these bimetallic chain compounds in order to verify
the validity of our method, a modified spin-wave theory, as well as
to understand typical features of 1-D ferrimagnets, and then proceed
to the above-mentioned trimeric copper phosphates and tetrameric
copper complex, whose Hamiltonians can be written as
\begin{mathletters}
   \begin{eqnarray}
   &&{\cal H}
     =J\sum_n
      (\mbox{\boldmath$S$}_n\cdot\mbox{\boldmath$s$}_n
      +\mbox{\boldmath$s$}_n\cdot\mbox{\boldmath$S$}_{n+1})\,,
   \label{E:Ha} \\
   &&{\cal H}
     =\sum_n
      \big[
       J_1
       (\mbox{\boldmath$s$}_{n,1}\cdot\mbox{\boldmath$s$}_{n,2}
       +\mbox{\boldmath$s$}_{n,2}\cdot\mbox{\boldmath$s$}_{n,3})
   \nonumber \\
   && \qquad\quad
      +J_2
       (\mbox{\boldmath$s$}_{n,2}\cdot\mbox{\boldmath$s$}_{n+1,1}
       +\mbox{\boldmath$s$}_{n,3}\cdot\mbox{\boldmath$s$}_{n+1,2})
      \big]\,,
   \label{E:Hb} \\
   &&{\cal H}
     =\sum_n
      \big[
       J_{\rm AF}
       (\mbox{\boldmath$s$}_{n,1}\cdot\mbox{\boldmath$s$}_{n,2}
       +\mbox{\boldmath$s$}_{n,2}\cdot\mbox{\boldmath$s$}_{n,3})
   \nonumber \\
   && \qquad\quad
      -J_{\rm F}
       (\mbox{\boldmath$s$}_{n,3}\cdot\mbox{\boldmath$s$}_{n,4}
       +\mbox{\boldmath$s$}_{n,4}\cdot\mbox{\boldmath$s$}_{n+1,1})
      \big]\,,
   \label{E:Hc}
   \end{eqnarray}
\end{mathletters}
respectively.
In the following, we take $N$ as the number of unit cells and set the
length of the unit cell to unity.

   Assuming the N\'eel-like order and introducing the bosonic
operators for the spin deviation in each sublattice, we can expand
the Hamiltonian (\ref{E:Ha}) as
\begin{equation}
   {\cal H}=E_{\rm class}+{\cal H}_0+{\cal H}_1+O(S^{-1})\,,
\end{equation}
where $E_{\rm class}=-2SsJN$ is the classical ground-state energy,
${\cal H}_0$ gives the free spin waves, and ${\cal H}_1$ describes
two-body interactions between them, which are, respectively, the
$O(S^2)$, $O(S^1)$, and $O(S^0)$ terms.
It may be an idea to diagonalize ${\cal H}_0$ and ${\cal H}_1$
simultaneously.
However, the resultant dispersion relations are gapped and thus
misread the low-temperature ferromagnetic features \cite{B3921}
inherent in 1-D ferrimagnets, such as the $T^{1/2}$-vanishing $C$
and the $T^{-2}$-diverging $\chi$.
Hence we propose another treatment \cite{Y11033} of the
up-to-$O(S^0)$ bosonic Hamiltonian.
${\cal H}_0$ is diagonalized as
\begin{equation}
   {\cal H}_0
     =E_0
     +J\sum_k
      \left(
       \omega_{k}^-\alpha_k^\dagger\alpha_k
      +\omega_{k}^+\beta_k^\dagger \beta_k
      \right)\,,
\end{equation}
where $E_0=J\sum_k[\omega_k-(S+s)]$ is the $O(S^1)$ quantum
correction to the ground-state energy, and $\alpha_k^\dagger$ and
$\beta_k^\dagger$ are the creation operators of the ferromagnetic and
antiferromagnetic spin waves of momentum $k$
whose dispersion relations are given by
$\omega_{k}^\pm=\omega_k\pm(S-s)$ with
$\omega_k=[(S-s)^2+4Ss\sin^2(k/2)]^{1/2}$.
Using the Wick theorem, ${\cal H}_1$ is rewritten as
%\begin{equation}
\begin{eqnarray}
   &&
   {\cal H}_1
    =E_1-J\sum_k
    \left(
    \delta\omega_k^-\alpha_k^\dagger\alpha_k
    +\delta\omega_k^+\beta_k^\dagger\beta_k
    \right)
   \nonumber\\
   &&\qquad
    +{\cal H}_{\rm irrel}+{\cal H}_{\rm resid}\,,
\end{eqnarray}
%\end{equation}
where the $O(S^0)$ correction to the ground-state energy and those to
the dispersions are given by
$E_1=-2JN
     [{\mit\Gamma}_1^2+{\mit\Gamma}_2^2
     +(\sqrt{S/s}+\sqrt{s/S}){\mit\Gamma}_1{\mit\Gamma}_2]$ and
$\delta\omega_k^\pm
 =2(S+s)({\mit\Gamma}_1/\omega_k)\sin^2(k/2)
 +({\mit\Gamma}_2/\sqrt{Ss})[\omega_k\pm (S-s)]$ with
${\mit\Gamma}_1=(2N)^{-1}\sum_k[(S+s)/\omega_k-1]$ and
${\mit\Gamma}_2=-N^{-1}\sum_k(\sqrt{Ss}/\omega_k)\cos^2(k/2)$,
while the irrelevant one-body terms
${\cal H}_{\rm irrel}
 =-J(S-s)^2({\mit\Gamma}_1/\sqrt{Ss})
  \sum_k[{\rm cos}(k/2)/{\omega_k}]
   (\alpha_k\beta_k+\alpha_k^\dagger\beta_k^\dagger)$
and the residual two-body interactions ${\cal H}_{\rm resid}$ are
both neglected so as to keep the ferromagnetic branch gapless.
This procedure may be recognized as the perturbational treatment of
${\cal H}_1$ to ${\cal H}_0$.
The resultant Hamiltonian is compactly represented as
\begin{equation}
   {\cal H}
     \simeq E_{\rm g}
    +J\sum_k
     \left(
      {\widetilde\omega}_k^- \alpha_k^\dagger \alpha_k
     +{\widetilde\omega}_k^+ \beta_k^\dagger  \beta_k
     \right)\,,
\end{equation}
with ${\widetilde\omega}_k^\pm=\omega_k^\pm-\delta\omega_k^\pm$ and
$E_{\rm g}=E_{\rm class}+E_0+E_1$.
\vspace*{2mm}
\begin{figure}
\centerline
{\mbox{\psfig{figure=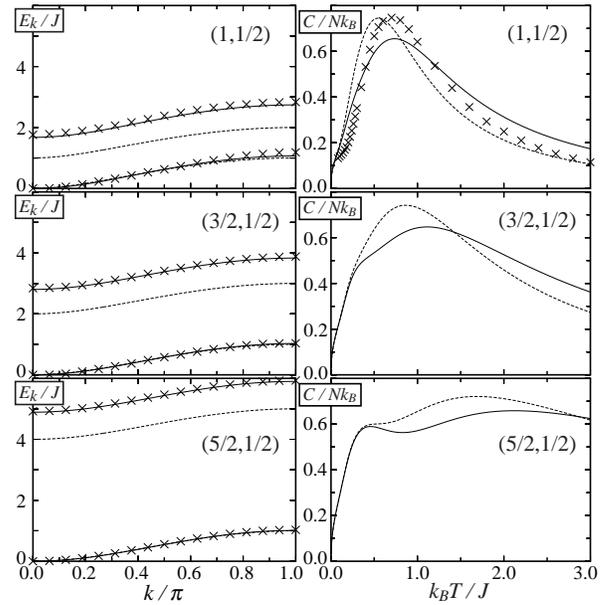,width=78mm,angle=0}}}
\vspace*{2mm}
\caption{Noninteracting (dashed lines) and interacting (solid lines)
         spin-wave calculations of the dispersion relations and the
         specific heat for the alternating-spin chains which are
         described by the Hamiltonian (1a) and Fig. 1(a).
         Corresponding quantum Monte Carlo calculations ($\times$)
         are also shown for comparison.}
\label{F:dimer}
\end{figure}

   In Fig. \ref{F:dimer}, the dispersion relations up to the order
$O(S^1)$ and $O(S^0)$, $\omega_k^\pm$ and ${\widetilde\omega}_k^\pm$,
are compared with the numerical findings obtained through a recently
developed quantum Monte Carlo scheme \cite{Y3348}.
We find two distinct branches of spin-wave excitations.
The gapless one is made of the elementary excitations reducing the
ground-state magnetization and is thus of ferromagnetic aspect, while
the gapped one, enhancing the ground-state magnetization, is of
antiferromagnetic aspect.
The ferromagnetic spin waves indeed exhibit a quadratic dispersion at
small momenta.
The $O(S^0)$ quantum correction has a significant effect on the
antiferromagnetic spin waves, whereas the ferromagnetic ones look
almost free from interaction.

   The core idea \cite{T233} of the so-called modified spin-wave
theory can be summarized as reliably describing thermodynamics even
in low dimensions by introducing a constraint on the magnetization.
We demonstrate its ferrimagnetic version \cite{Y14008} for the
alternating-spin chains.
Constraining the total magnetization to be zero, Takahashi
\cite{T168} obtained an excellent description of the low-temperature
thermodynamics of 1- and 2-D Heisenberg ferromagnets.
His idea that the thermal spin deviation should be equal to the
ground-state magnetization may be replaced by
\begin{equation}
   (S+s)N=(S+s)\sum_k(n_k^-+n_k^+)/\omega_k\,,
   \label{E:const}
\end{equation}
for our ferrimagnets, where
$n_k^\pm=\sum_{n^-,n^+}n^\pm P_k(n^-,n^+)$ with $P_k(n^-,n^+)$
being the probability of $n^-$ ferromagnetic and $n^+$
antiferromagnetic spin waves appearing in the $k$-momentum state.
Equation (\ref{E:const}) claims that the thermal fluctuation should
cancel the N\'eel-state {\it staggered} magnetization instead of the
{\it uniform} one, in response to the replacement of the
ferromagnetic exchange coupling by the antiferromagnetic one.
Minimizing the free energy
%\begin{equation}
\begin{eqnarray}
   &&
   F=E_{\rm g}
    +\sum_k\sum_{\sigma=\pm} n_k^\sigma\widetilde\omega_k^\sigma
   \nonumber \\
   &&\qquad 
   +k_{\rm B}T
     \sum_k\sum_{n^-,n^+}P_k(n^-,n^+){\rm ln}P_k(n^-,n^+)\,,
\end{eqnarray}
%\end{equation}
with respect to $P_k$ at each $k$ under the condition (\ref{E:const})
as well as the trivial constraints $\sum_{n^-,n^+}P_k(n^-,n^+)=1$,
we obtain the free and internal energies at thermal equilibrium as
$F=E_{\rm g}+\mu(S-s)N
  -k_{\rm B}T\sum_{k}\sum_{\sigma=\pm}
   {\rm ln}(1+\bar n^\sigma_k)$ and
$U=E_{\rm g}+J\sum_k\sum_{\sigma=\pm}
   \bar n_k^\sigma\widetilde \omega_k^\sigma$,
where the optimum distributions are given by
$\bar n_k^\pm
 =\{{\mbox e}^{[J{\widetilde\omega}_k^\pm
    -\mu(S+s)/\omega_k]/k_{\rm B}T}-1\}^{-1}$
with a Lagrange multiplier $\mu$ obtained through Eq.
(\ref{E:const}).
The thus-obtained specific heat, together with its interaction-free
version, is shown in Fig. \ref{F:dimer}.
In the case of $(S,s)=(1,\frac{1}{2})$, we compare the present
results with the quantum Monte Carlo estimates.
Considering that the conventional antiferromagnetic spin-wave theory
does not work at all for the 1-D thermodynamics, the present
calculations surprisingly well describe the overall temperature
dependence, including the $\sqrt{T}$ initial behavior and the
Schottky-type peak.
When we employ the interacting spin waves, the most significant
improvement is the correction of the peak position.
Although they somewhat underestimate the peak height, the
Schottky-type anomaly is then correctly located, which is essential
to reproduce the overall structure.
The specific heat may be double-peaked provided the antiferromagnetic
gap is much larger than the ferromagnetic band width, which is
expressed in terms of the interacting spin waves as
$\widetilde\omega_{k=0}^+ \gg
 \widetilde\omega_{k=\pi}^- - \widetilde\omega_{k=0}^-$.
This condition is satisfied for $S\gg 2s$ and the ferrimagnetic chain
of $(S,s)=(\frac{5}{2},\frac{1}{2})$ indeed exhibits a double-peaked
specific heat.
In the following, we argue the other mechanism for a varied
temperature dependence, which is of topological origin and may thus
be valid for homometallic 1-D ferrimagnets.
Considering the poor convergence of quantum Monte Carlo calculations
at low temperatures, the present modified spin-wave scheme is one of
the most reliable and feasible approach.
%\vspace*{-2mm}
\begin{figure}
\centerline
{\mbox{\psfig{figure=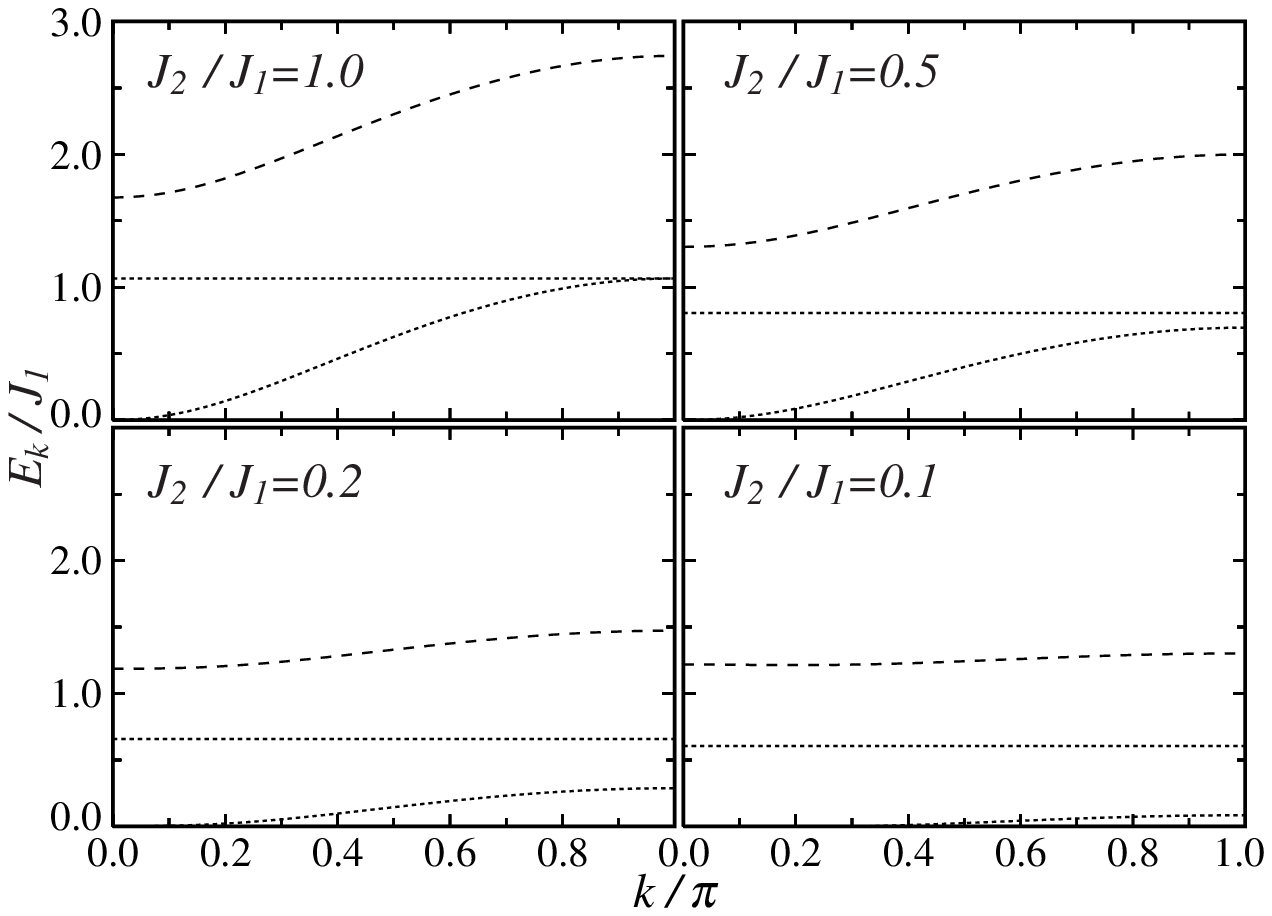,width=78mm,angle=0}}}
\vspace*{2mm}
\caption{Interacting spin-wave calculations of the dispersion
         relations for the trimeric chains which are described by
         the Hamiltonian (1b) and Fig. 1(b), where the ferromagnetic
         and antiferromagnetic excitations are distinguishably shown
         by dotted and dashed lines, respectively.}
\label{F:trimerE}
\vspace*{6mm}
\centerline
{\mbox{\psfig{figure=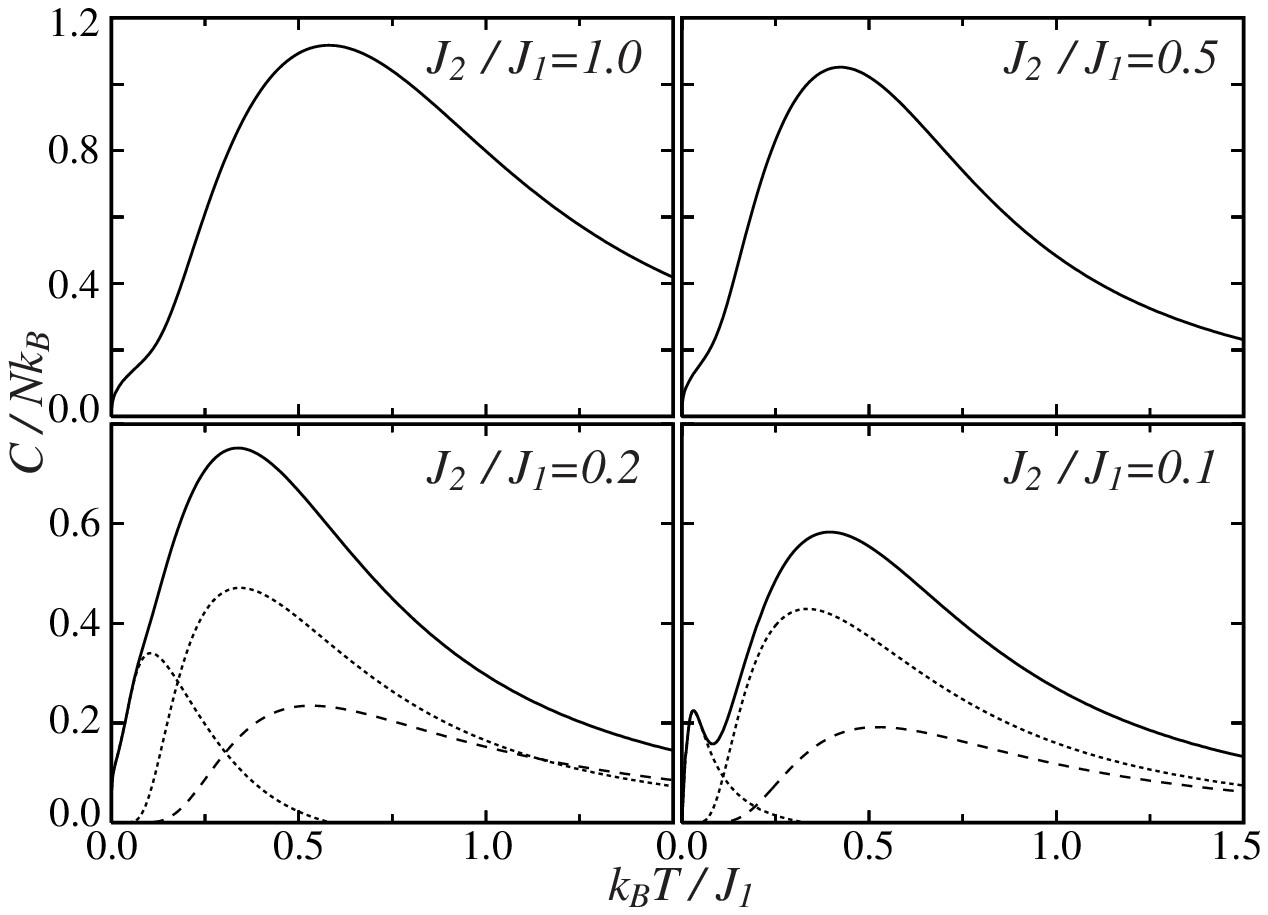,width=78mm,angle=0}}}
\vspace*{2mm}
\caption{Interacting spin-wave calculations of the specific heat for
         the trimeric chains which are described by the Hamiltonian
         (1b) and Fig. 1(b), where individual contributions of the
         ferromagnetic and antiferromagnetic spin waves are also
         shown by dotted and dashed lines, respectively, the sum of
         which is equal to the total drawn by solid lines.}
\label{F:trimerC}
\end{figure}
\vspace*{-2mm}

   The dispersion relations and the resultant specific heat of the
trimeric intertwining chains are shown in Figs. \ref{F:trimerE} and
\ref{F:trimerC}, respectively.
The system is analogous to the alternating-spin chain of
$(S,s)=(1,\frac{1}{2})$ but displays an additional flat band which is
gapped but of ferromagnetic aspect.
With decreasing $J_2$ in comparison with $J_1$, a second anomaly
appears at low temperatures.
The present tool advantageously enables us to observe each
contribution of the distinct excitation bands.
We find that due to the existence of the mid band, the double-peak
structure is limited to rather small ratios $J_2/J_1$.
In the previous experiments \cite{D83}, the minimum related to the
second anomaly was indeed observed, but the measured temperatures
($1.8\sim 15$ K) were not low enough to estimate the maximum of the
bump.
A significant increase and the following Schottky-type main peak at
higher temperatures were not explicitly shown either, where the
magnetic and lattice contributions should carefully be separated.
Our calculations fully motivate further measurements \cite{A186} and
make possible detailed analyses of them.
%\vspace*{-2mm}
\begin{figure}
\centerline
{\mbox{\psfig{figure=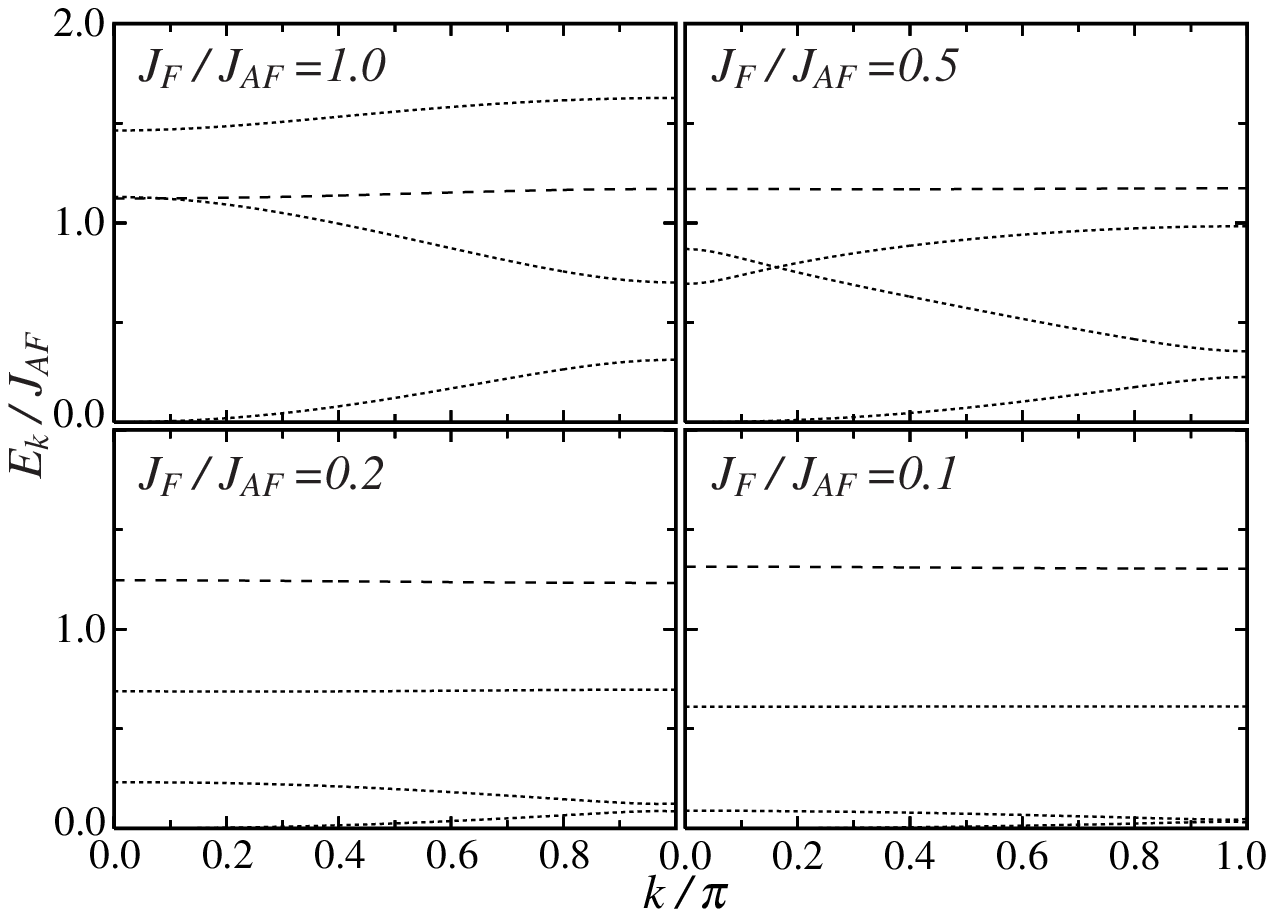,width=78mm,angle=0}}}
\vspace*{2mm}
\caption{Interacting spin-wave calculations of the dispersion
         relations for the tetrameric chains which are described by
         the Hamiltonian (1c) and Fig. 1(c), where the ferromagnetic
         and antiferromagnetic excitations are distinguishably shown
         by dotted and dashed lines, respectively.}
\label{F:tetramerE}
\vspace*{6mm}
\centerline
{\mbox{\psfig{figure=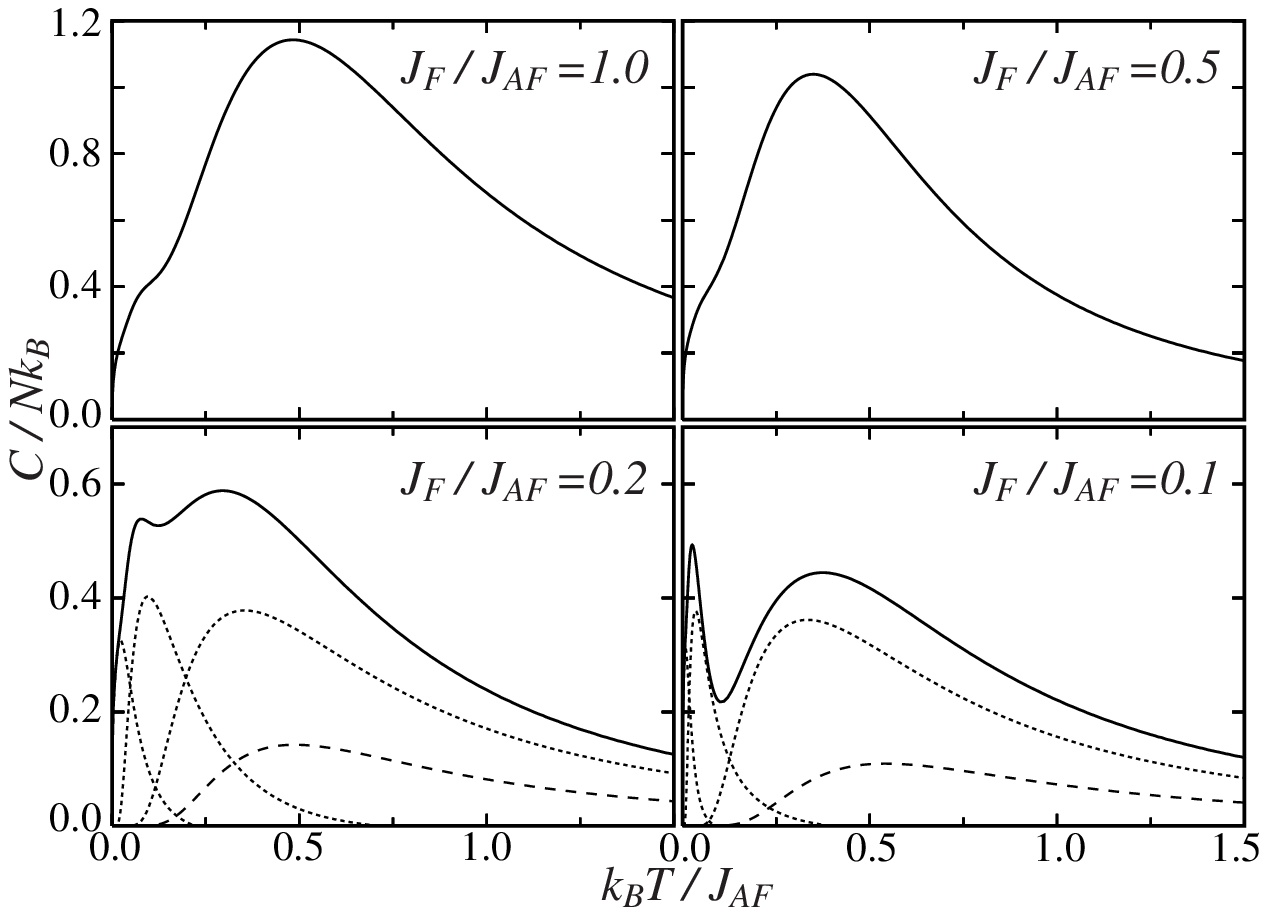,width=78mm,angle=0}}}
\vspace*{2mm}
\caption{Interacting spin-wave calculations of the specific heat for
         the tetrameric chains which are described by the Hamiltonian
         (1c) and Fig. 1(c), where individual contributions of the
         ferromagnetic and antiferromagnetic spin waves are also
         shown by dotted and dashed lines, respectively, the sum of
         which is equal to the total drawn by solid lines.}
\label{F:tetramerC}
\end{figure}

   The tetrameric bond-alternating chains are discussed in Figs.
\ref{F:tetramerE} and \ref{F:tetramerC}.
We again find a double-peaked specific heat of topological origin.
The lower two bands construct the low-temperature bump, while the
upper two contribute to the main peak at mid temperatures.
It is the double contribution that makes the second peak much more
noticeable in the tetrameric chains than in the trimeric chains.
The susceptibility measurements \cite{H30,A4466} were well
interpreted by setting $J_{\rm F}/J_{\rm AF}$ to $0.5$.
However, such parameters neither induce any detectable second peak of
the specific heat within the Hamiltonian (\ref{E:Hc}) nor fit the
recent observations \cite{H} detecting the second anomaly around
$0.5$ K.
We have tried the opposite parametrizations
$J_{\rm AF}<J_{\rm F}$.
However, the third band lying much closer to the lower two makes the
low-temperature anomaly much more conspicuous and suppress the
Schottky-type peak, ending up with complete discrepancy with the
observations.
Further measurements and more extensive analyses, for instance,
taking account of exchange anisotropy, are necessary for the total
understanding.

   The trimeric and tetrameric chain materials have been measured by
separate groups under their respective motivations.
Interestingly, however, the unique temperature dependences of their
specific heat potentially originate in the same mechanism.
We have revealed {\it the intrinsic double-peak structure of the
specific heat in low-D ferrimagnets of topological origin}.
From this point of view, homometallic or organic ferrimagnets should
further be synthesized and measured, where the double-peaked specific
heat may generically be observed.

   This work was supported by the Japanese Ministry of Education,
Science, and Culture and by the Sumitomo Foundation.
The numerical calculation was done using the facility of the
Supercomputer Center, Institute for Solid State Physics, University
of Tokyo.

\widetext

\begin{references}

\bibitem{K89}
   O. Kahn,
      Struct. Bonding (Berlin) {\bf 68}, 89 (1987);
   O. Kahn, Y. Pei, and Y. Journaux,
      in {\it Inorganic Materials},
      edited by D. W. Bruce and D. O'Hare (Wiley, New York, 1995),
      p. 95.

\bibitem{Y1024}
   S. Yamamoto,
      Phys. Rev. B {\bf 59}, 1024 (1999);
   S. Yamamoto, T. Fukui, and T. Sakai,
      Eur. Phys. J. B. {\bf 15} (2000) 211.


\bibitem{O1984}
   M. Oshikawa, M. Yamanaka, and I. Affleck,
      Phys. Rev. Lett. {\bf 78}, 1984 (1997).

\bibitem{K1762}
   T. Kuramoto,
      J. Phys. Soc. Jpn. {\bf 67}, 1762 (1998);
                         {\bf 68}, 1813 (1999).

\bibitem{S4053}
   T. Sakai and S. Yamamoto,
      Phys. Rev. B {\bf 60}, 4053 (1999);
      J. Phys.: Condens. Matter {\bf 12}, 9787 (2000).

\bibitem{L11725}
   A. Langari and M. A. Martin-Delgado,
      Phys. Rev. B {\bf 62}, 11725 (2000).

\bibitem{Y3795}
   S. Yamamoto and T. Sakai,
      Phys. Rev. {\bf 62}, 3795 (2000).

\bibitem{D409}
   J. Darriet and L. P. Regnault,
      Solid State Commun. {\bf 86}, 409 (1993).

\bibitem{D1857}
   J. F. DiTusa, S.-W. Cheong, J.-H. Park, G. Aeppli, C. Broholm,
   and C. T. Chen,
      Phys. Rev. Lett. {\bf 73}, 1857 (1994).

\bibitem{Z6437}
   A. Zheludev, J. M. Tranquada, T. Vogt, and D. J. Buttrey,
      Phys. Rev. {\bf 54}, 6437 (1996);
                 {\bf 54}, 7210 (1996).

\bibitem{Y11516}
   T. Yokoo, A. Zheludev, M. Nakamura, and J. Akimitsu,
      Phys. Rev. {\bf 55}, 11516 (1997);
   T. Yokoo, S. Raymond, A. Zheludev, S. Maslov, E. Ressouche,
   I. Zaliznyak, R. Erwin, M. Nakamura, and J. Akimitsu,
      {\it ibid.}, {\bf 58}, 14424 (1998).

\bibitem{T15189}
   Y. Takushima, A. Koga, and N. Kawakami,
      Phys. Rev. {\bf 61}, 15189 (2000).

\bibitem{F433}
   N. Fujiwara and M. Hagiwara,
      Solid State Commun. {\bf 113}, 433 (2000).

\bibitem{Y842}
   S. Yamamoto,
      Phys. Rev. {\bf 61}, R842 (2000);
      Phys. Lett. A {\bf 265}, 139 (2000);
      J. Phys. Soc. Jpn. {\bf 69}, 2324 (2000).

\bibitem{D83}
   M. Drillon, M. Belaiche, P. Legoll, J. Aride, A. Boukhari, and
   A. Moqine,
      J. Magn. Magn. Mater. {\bf 128}, 83 (1993).

\bibitem{H}
   M. Hagiwara and H. Katori,
      private communication.

\bibitem{M5908}
   K. Maisinger, U. Schollw\"ock, S. Brehmer, H.-J. Mikeska, and
   S. Yamamoto,
      Phys. Rev. B {\bf 58}, R5908 (1998).

\bibitem{A186}
   Y. Ajiro, T. Asano, K. Nakaya, M. Mekata, K. Ohyama, Y. Yamaguchi,
   Y. Koike, Y. Morii, K. Kamishima, H. Katori, and T. Goto,
      J. Phys. Soc. Jpn. Suppl. A {\bf 70}, 186 (2001).

\bibitem{H30}
   M. Hagiwara, Y. Narumi, K. Minami, and K. Kindo,
      Physica B {\bf 294}-{\bf 295}, 30 (2001).

\bibitem{K782}
   O. Kahn, Y. Pei, M. Verdaguer, J.-P. Renard, and J. Sletten,
      J. Am. Chem. Soc. {\bf 110}, 782 (1988);
   P. J. van Koningsbruggen, O. Kahn, K. Nakatani, Y. Pei,
   J.-P. Renard, M. Drillon, and P. Legoll,
      Inorg. Chem. {\bf 29}, 3325 (1990).

\bibitem{B3921}
   S. Brehmer, H.-J. Mikeska, and S. Yamamoto,
      J. Phys.: Condens. Matter {\bf 9}, 3921 (1997);
   S. Yamamoto, S. Brehmer, and H.-J. Mikeska,
      Phys. Rev. B {\bf 57}, 13610 (1998).

\bibitem{Y11033}
   S. Yamamoto, T. Fukui, K. Maisinger, and U. Schollw\"ock,
      J. Phys.: Condens. Matter {\bf 10}, 11033 (1998).

\bibitem{Y3348}
   S. Yamamoto,
      Phys. Rev. Lett. {\bf 75}, 3348 (1995);
   S. Yamamoto and S. Miyasita,
      Phys. Lett. A {\bf 235}, 545 (1997).

\bibitem{T233}
   M. Takahashi,
      Prog. Theor. Phys. Suppl. {\bf 87}, 233 (1986).

\bibitem{Y14008}
   S. Yamamoto and T. Fukui,
      Phys. Rev. B {\bf 57}, 14008 (1998).

\bibitem{T168}
   M. Takahashi,
      Phys. Rev. Lett. {\bf 58}, 168 (1987).

\bibitem{A4466}
   A. Escuer, R. Vicente, M. S. E. Fallah, M. A. S. Goher, and
   F. A. Mautner,
      Inorg. Chem. {\bf 37}, 4466 (1998).


\end{references}
\end{document}